\documentclass[aps,prb,twocolumn,superscriptaddress, show pacs]{revtex4-1}
\usepackage{bm, amsmath}
\usepackage{graphicx}
\usepackage{color}
\usepackage{epstopdf}
\usepackage{graphicx}
\usepackage{dcolumn}
\usepackage{bm}
\usepackage{subfigure}
\usepackage{placeins}

\begin{document}

\title{Internal structure of hexagonal skyrmion lattices in cubic helimagnets}

\author{D. McGrouther} 
\affiliation{School  of  Physics  and  Astronomy, University  of  Glasgow,  Glasgow,  UK,  G12  8QQ}
   
\author{R. J. Lamb} 
\affiliation{School  of  Physics  and  Astronomy, University  of  Glasgow,  Glasgow,  UK,  G12  8QQ}
 
\author{M. Krajnak} 
\affiliation{School  of  Physics  and  Astronomy, University  of  Glasgow,  Glasgow,  UK,  G12  8QQ}

\author{S. McFadzean} 
\affiliation{School  of  Physics  and  Astronomy, University  of  Glasgow,  Glasgow,  UK,  G12  8QQ}

\author{S. McVitie} 
\affiliation{School  of  Physics  and  Astronomy, University  of  Glasgow,  Glasgow,  UK,  G12  8QQ}
 
\author{R. L. Stamps} 
\affiliation{School  of  Physics  and  Astronomy, University  of  Glasgow,  Glasgow,  UK,  G12  8QQ}
 
\author{A.~O.~Leonov}
\affiliation{Center for Chiral Science, Hiroshima University, Higashi-Hiroshima, 
Hiroshima 739-8526, Japan}
\affiliation{IFW Dresden, Postfach 270016, D-01171 Dresden, Germany}   
\affiliation{Zernike Institute for Advanced Materials, University of Groningen, 
Groningen, 9700AB, The Netherlands}

\author{A.~N.~Bogdanov}
\affiliation{Center for Chiral Science, Hiroshima University, Higashi-Hiroshima, 
Hiroshima 739-8526, Japan}
\affiliation{IFW Dresden, Postfach 270016, D-01171 Dresden, Germany}
 
\author{Y. Togawa} 
\affiliation{School  of  Physics  and  Astronomy, University  of  Glasgow,  Glasgow,  UK,  G12  8QQ}    
\affiliation{Center for Chiral Science, Hiroshima University, Higashi-Hiroshima, 
Hiroshima 739-8526, Japan}
\affiliation{Osaka Prefecture University, 1-2 Gakuencho, Sakai, Osaka 599-8570, Japan} 
\affiliation{JST, PRESTO, 4-1-8 Honcho Kawaguchi, Saitama 333-0012, Japan}

\date{\today}

\begin{abstract}
{We report the most precise observations to date concerning the spin structure of magnetic skyrmions in a nanowedge specimen of cubic B20 structured FeGe. Enabled by our development of advanced Differential Phase Contrast (DPC) imaging (in a scanning transmission electron microscope (STEM)) we have obtained high spatial resolution quantitative measurements of skyrmion internal spin profile. For hexagonal skyrmion lattice cells, stabilised by an out-plane applied magnetic field, mapping of the in-plane component of magnetic induction has revealed precise spin profiles and that the internal structure possesses intrinsic six-fold symmetry. With increasing field strength, the diameter of skyrmion cores was measured to decrease and accompanied by a non-linear variation of the lattice periodicity. Variations in structure for individual skyrmions across an area of the lattice were also studied utilising a new increased sensitivity DPC detection scheme and a variety of symmetry lowering distortions were observed. To provide insight into fundamental energetics we have constructed a phenomenological model, with which our experimental observations of spin profiles and field induced core diameter variation are in good agreement with predicted structure in the middle of the nanowedge crystal. In the vicinity of the crystal surfaces, our model predicts the existence of in-plane twisting distortions which our current experimental observations were not sensitive to. As an alternative to the requirement for as yet unidentified sources of magnetic anisotropy, we demonstrate that surface states could provide the energetic stabilisation needed for predomination over the conical magnetic phase.}
\end{abstract}

         
\maketitle

\vspace{5mm}

\section{Introduction}

It was established theoretically quite some time ago, that in magnetic systems with broken inversion symmetry, the \textit{Dzyaloshinskii-Moriya} (DM) exchange interactions stabilize long-period modulations of the magnetization vector propagating with a fixed rotation sense along one (\textit{helices}) \cite{Dz64} or two (\textit{skyrmions}) \cite{JETP89,JMMM94} spatial directions.

Since the year 1976, chiral helical modulations have been identified by magnetization measurements and polarized small-angle neutron scattering techniques in several classes  of bulk noncentrosymmetric magnetic crystals including an extending group of cubic helimagnets with B20-type  structure (space group $P$2$_1$3) \cite{Ishikawa76,Lebech89}. Specific magnetic anomalies reported to occur near the ordering temperature of bulk cubic helimagnets indicate the existence of multidimensional chiral modulations (see e.g. \cite{Muehlbauer09,Pappas09,Wilhelm11}) and are in accordance with theoretical predictions of skyrmion textures \cite{Roessler06,Wilhelm12}.
Recently, the fabrication of high-quality thin films of noncentrosymmetric cubic and uniaxial ferromagnets \cite{Uchida06,Togawa12,Yu10,Yu11,Yu15,Wilson12,Yokouchi15} and nanolayers of common magnetic metals with surface/interface DM interactions \cite{Romming13,Romming15} have provided a material basis for detailed investigations of chiral modulations and possibilities for applications in spintronic devices \cite{Kiselev11,Fert13}.  
Despite such extensive focus, the precise spin-structure of skyrmions and the basis for their energetic predominance over the helical and conical states \cite{Butenko10,Wilson14}, which also arise from competition between symmetric ferromagnetic exchange and antisymmetric DM interaction, has not yet been fully explained \cite{JMMM94,Butenko10, Muhlbauer2009,Meynell14}. More recently a number of image based observations, utilising image reconstruction based on the transport of intensity equation (TIE), from the Fresnel mode of Lorentz transmission electron microscopy (TEM) or magnetic force microscopy \cite{Milde13} have provided insight into general skyrmion form, lattice arrangement and behaviour. Holographic TEM imaging has been applied to lattice skyrmions in (Fe,Co)Si  where the reconstruction obtained provided higher, but not precisely quoted, spatial resolution characterisation \cite{Park14}. The resulting interpretation was that lattice skyrmions possessed cylindrical symmetry and showed no variation in magnetic structure through the thickness of the nano-wedge sample. 
In Ref. \cite{Romming15} spin-polarized scanning tunneling microscopy (SP-STM) has been applied to resolve the internal structure of isolated skyrmions in nanolayers PdFe/Ir (111) with interface induced DM interactions. These skyrmions are an order of magnitude smaller and possess spin structures distinct from lattice skyrmions observed in bulk cubic heli-magnets.

In this article, we report results obtained using a novel high spatial resolution imaging method capable of providing quantitative measurements of the magnetic induction distribution in modulated states of different classes of noncentrosymmetric magnets. Using differential phase contrast (DPC) imaging, performed in a scanning transmission electron microscope (STEM), we reveal the internal magnetic spin profile of lattice skyrmions in a freestanding nano-wedge of FeGe at T=250K. Stabilized by the influence of an applied magnetic field, our imaging shows that the average lattice skyrmion possesses intrinsic six-fold symmetry and profile measurements reveal that the width of the skyrmion core decreases as the field strength is increased, accompanied by changes in the lattice periodicity.  
Furthermore, through utilising an enhanced detection scheme we show that, although the arrangement of lattice skyrmions is highly periodic, there is a significant level of individual structure variation which for some skyrmions results in their spin profile symmetry being reduced. This observation documents a more subtle alteration than recent works which have highlighted distortions to the skyrmion lattice through Fresnel imaging\cite{Rajeswari15,shibata_naturenano} and to skyrmions themselves, also by DPC imaging \cite{matsumoto_scienceadvances}, at a crystal grain boundary.

Within a standard phenomenological model  \cite{Dz64,Bak80} we derive rigorous solutions for skyrmion lattice cells in thin layers of cubic helimagnets at different applied fields. We also predict that the structure of the skyrmion lattice includes regions of twisting of the in-plane magnetisation close to the material surfaces in which it is confined and that these surface states lead to energetic stabilisation of the skyrmion lattice over the conical state. Finally we consider our experimentally observed evolution of the skyrmion lattice structure and find consistency with our presented theoretical model.

\begin{figure*}
\includegraphics[width=16cm]{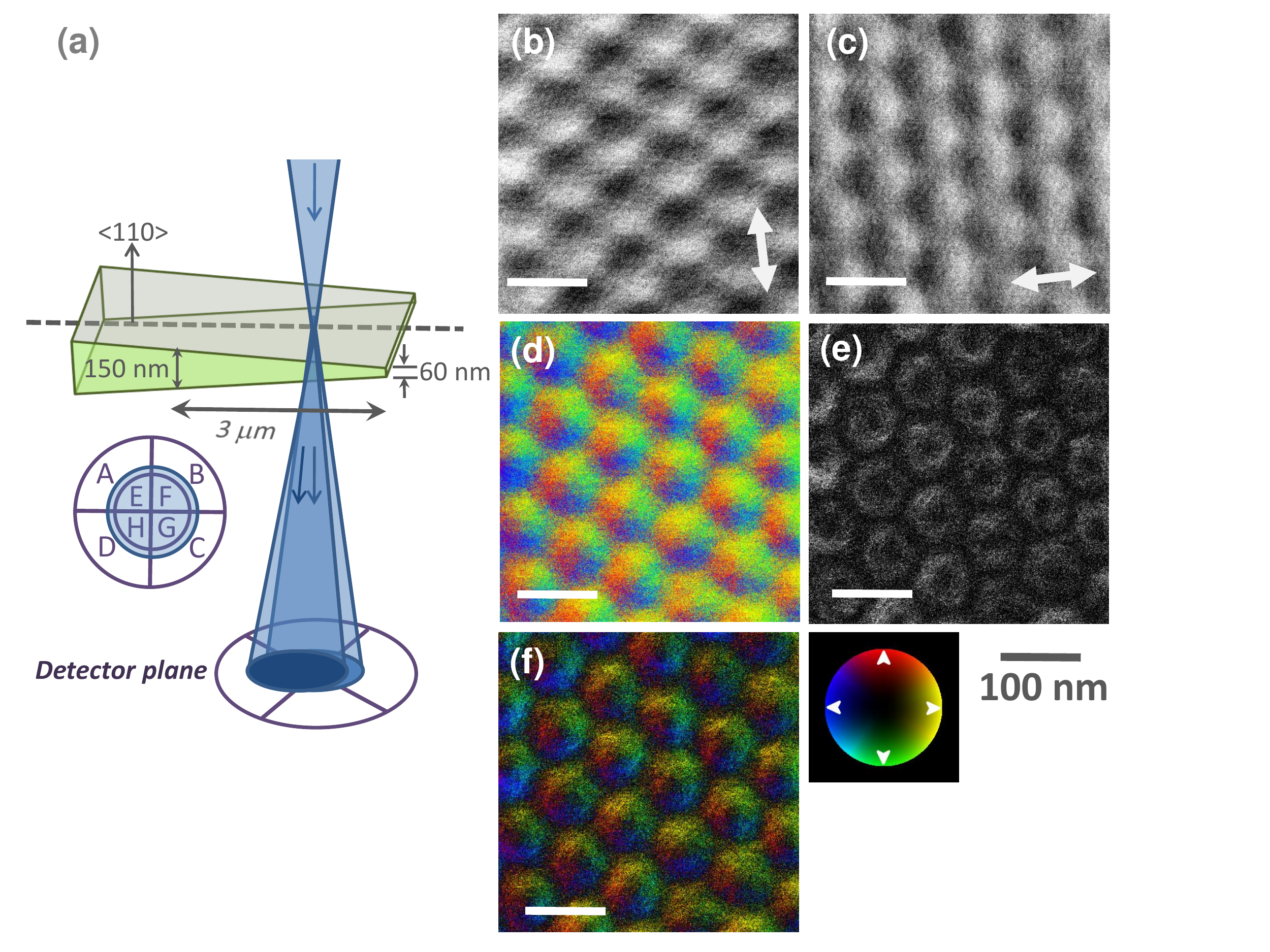}
\caption{(a) Schematic diagram of the specimen geometry and principle of the DPC technique. (b) $\&$ (c) Orthogonal DPC component images with directions of sensitivity indicated by double headed arrows. (d) Colour respresentation of in-plane induction directions. (e) Total magnitude of in-plane components. (f) Colour vector representation of in-plane directions from combining (d) $\&$ (e) }
\label{fig:dpc_explain_skyrmions}
\end{figure*}

\section{Experimental measurement of structure}
Benefitting from developments in aberration correction for electron optics we have recently improved the range of sensitivity and spatial resolution, to better than $1 nm$, available from the Differential Phase Contrast (DPC) imaging mode \cite{mcvitie_ultramicroscopy, togawa15, krajnak_pixdpc}. For the purpose of imaging skyrmions in B20 structured FeGe, a wedge shaped specimen with $<110>$ normal was extracted from a vapour grown bulk single crystal sample (details of preparation and characterisation given in Supplementary Information). The thickness of the wedge increased linearly from $60-150 nm$ over a distance of $3 \mu m$ from it's apex and was thin enough to be suitably transmitting for 200keV electrons, figure \ref{fig:dpc_explain_skyrmions}. For electrons at normal incidence the beam is only sensitive to the in-plane magnetic induction components of the skyrmions as dictated by understanding based on the wave-optical Aharanov-Bohm effect \cite{chapman_jphysd1984}. As the focused electron beam possesses much smaller diameter than the period of the in-plane magnetisation curling, for each point in a raster scan, the phase shift of the traversing electron waves manifests itself at the detector plane as an angular deflection of the transmitted electron cone, as shown in figure \ref{fig:dpc_explain_skyrmions}(a). For this work we have chosen to employ an increased probe size, half-width $3 nm$, in order to obtain a reduced probe semi-convergence angle of $415 \mu radians$ and therefore greater sensitivity to $\mu radian$ angular displacements.

Direct viewing of the angular displacements is achieved by forming orthogonal difference images (e.g. A-C and B-D) and constitutes a differentiation of the electron wave-phase gradients induced by the sample. Figures \ref{fig:dpc_explain_skyrmions}(b) $\&$ (c) show the orthogonal DPC difference images obtained from a relatively flat region of the specimen containing lattice skyrmions with an applied field (supplied by the partially excited objective lens) of strength $H_{appl} = 796 Oe$. Performing DPC imaging with an annular, rather than solid, detector provides a means by which to suppress unwanted differential contrast from non-magnetic origins \cite{Chapman_MDPC}. In this case signals from the inner segments E,F,G,H were used to subtract minor differential contrast variations from diffraction by weak bend contours. As can be observed in figures \ref{fig:dpc_explain_skyrmions}(b) $\&$ (c) both show strong contrast with periodic variation of the grey level. Further direct maps of the variation of direction and strength of the in-plane magnetic induction can be revealed by mathematically combining the information from the orthogonal component images figures \ref{fig:dpc_explain_skyrmions}(b) $\&$ (c). An arc-tangent operation yields the in-plane induction directions which are mapped to a colourwheel. Figure \ref{fig:dpc_explain_skyrmions}(d) shows the periodic in-plane rotation associated with each of 15 full skyrmions that can be seen, in this case all possessing clockwise chirality as dictated by the single sign of the DMI constant. The overall magntiude of the in-plane moments are obtained through quadratic summation of the difference images (e.g. $\sqrt{{(A-C)}^{2} + {(B-D)}^{2}}$), figure \ref{fig:dpc_explain_skyrmions}(e), and show that each skyrmion is characterised by a ring of intensity, corresponding to where the moments gain an in-plane component as they twist from one out-of-plane orientation at the periphery of the skyrmion to the opposite out-plane-orientation at it's core. Combining information from both figures \ref{fig:dpc_explain_skyrmions}(d) $\&$ (e) yields a quantitative in-plane vector image of the skyrmion lattice, figure \ref{fig:dpc_explain_skyrmions}(f). This figure bears close resemblance to vector maps produced from TIE-based reconstructions already published. However, in this case the intrinsic spatial resolution, at $3 nm$, is higher and thus the image contains contributions from a much wider range of spatial frequencies. 
\begin{figure*}
\includegraphics[width=16cm]{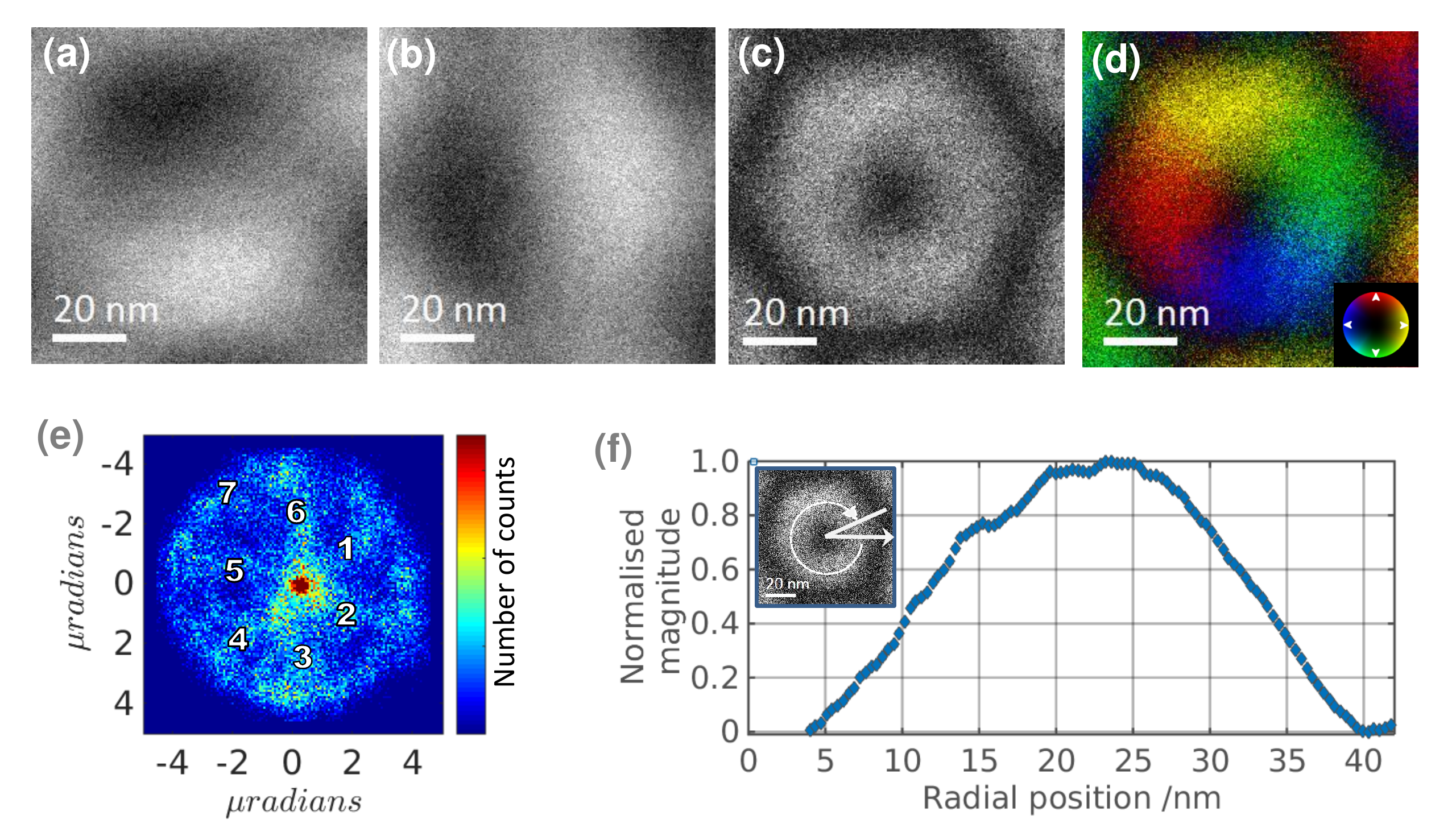}
\caption{ (a) $\&$ (b) DPC component images of an average lattice skyrmion and resulting (c) magnitude $\&$ (d) colour vector in-plane maps. (e) Bi-variate histogram analysis showing the distribution of magnitude and direction information from (a) $\&$ (b). (f) Radially averaged intensity profile from the magnitude image (c) produced by circularly sweeping the line indicated in the inset magnitude figure.
\label{fig:dpc_sk_averaged}}
\end{figure*}
In order to reduce the influence from statistical noise and natural variations in the lattice skyrmion structure, averaged DPC difference images, figures \ref{fig:dpc_sk_averaged}(a),(b) were produced by calculating mean intensity maps over a number, $n=15$, of lattice skyrmions from figures \ref{fig:dpc_explain_skyrmions}(b),(c). The averaged difference images were then combined to produce maps of the in-plane magnitude, figure \ref{fig:dpc_sk_averaged}(c) and the in-plane colour vectors, figure \ref{fig:dpc_sk_averaged}(d). Taken together figures \ref{fig:dpc_sk_averaged} (c)$\&$(d) yield a clear view that the average lattice skyrmion possesses a hexagonal form. From the averaged DPC difference images figures \ref{fig:dpc_sk_averaged}(b),(c) a bivariate histogram analysis was performed,figure \ref{fig:dpc_sk_averaged}(e), allowing visualisation of the quantitative distribution of electron beam deflection, $\beta$, from across the skyrmion structure. From figure \ref{fig:dpc_sk_averaged}(e) it can be observed that the histogram is composed of six spokes, numbered 1-6, and an outer ring, numbered 7. The latter relates to the strongest deflections experienced by beam and were measured at $\pm 4 \mu radians$, corresponding to regions where the skyrmion is magnetised completely in-plane. The magnitude of maximum deflection angle $(\beta = {e \lambda B_{S}t} /{h})$ is equivalent to a magnetic induction thickness product, $B_{S}t = 13.3 T\cdot nm$. The region of the specimen in which the images were acquired was measured to have thickness in the range $t=60-80 nm$ and thus the strength of the induction can be inferred to be $B=0.2 T$. As the electron beam is only sensitive to in-plane components of induction, in the absence of strong domain wall features with associated magnetisation divergence acting as sources of stray magnetic field,it is justifiable to assume that $\beta$ is a measurement of the strength of the in-plane magnetisation. Our value for the magnetic induction agrees well with measurements of magnetisation performed for bulk FeGe. Lundgren et al. \cite{lundgren_ms_measurement} measured a magnetic moment of $m_S=1\mu_{B}$ per Fe atom which translates to a saturation induction $B_S=0.411T$. From a fitted Brillouin function, at $T=250K$ they also measured magnetisation to be reduced by $50\%$ , yielding $B=0.207T$. Returning to figure \ref{fig:dpc_sk_averaged}(e), the bivariate histogram provides a strong depiction of intrinsic symmetry within the skyrmion through the presence of six spokes, numbered 1-6,  projecting inwards towards the histogram centre and the hexagonal appearance of the outer ring, numbered 7.  Figure \ref{fig:dpc_sk_averaged}(f) plots the radially averaged intensity profile produced by sweeping the line, indicated in figure \ref{fig:dpc_sk_averaged}(c), through a full rotation about the skyrmion core. From figure \ref{fig:dpc_sk_averaged}(f) it can be seen that the average skyrmion has a measured radius of 40 $nm$. Taking together all of our imaging and analyses in figure \ref{fig:dpc_sk_averaged} we deduce that in common with their lattice symmetry, the average lattice skyrmion possesses a hexagonally symmetric internal structure in this case. Previously it has been reported that purely cylindrical rotation was observed by electron holography in FeCoSi\cite{Park14}, our measurements indicate a more subtle variation in the FeGe system.

\begin{figure*}
\includegraphics[width=18cm]{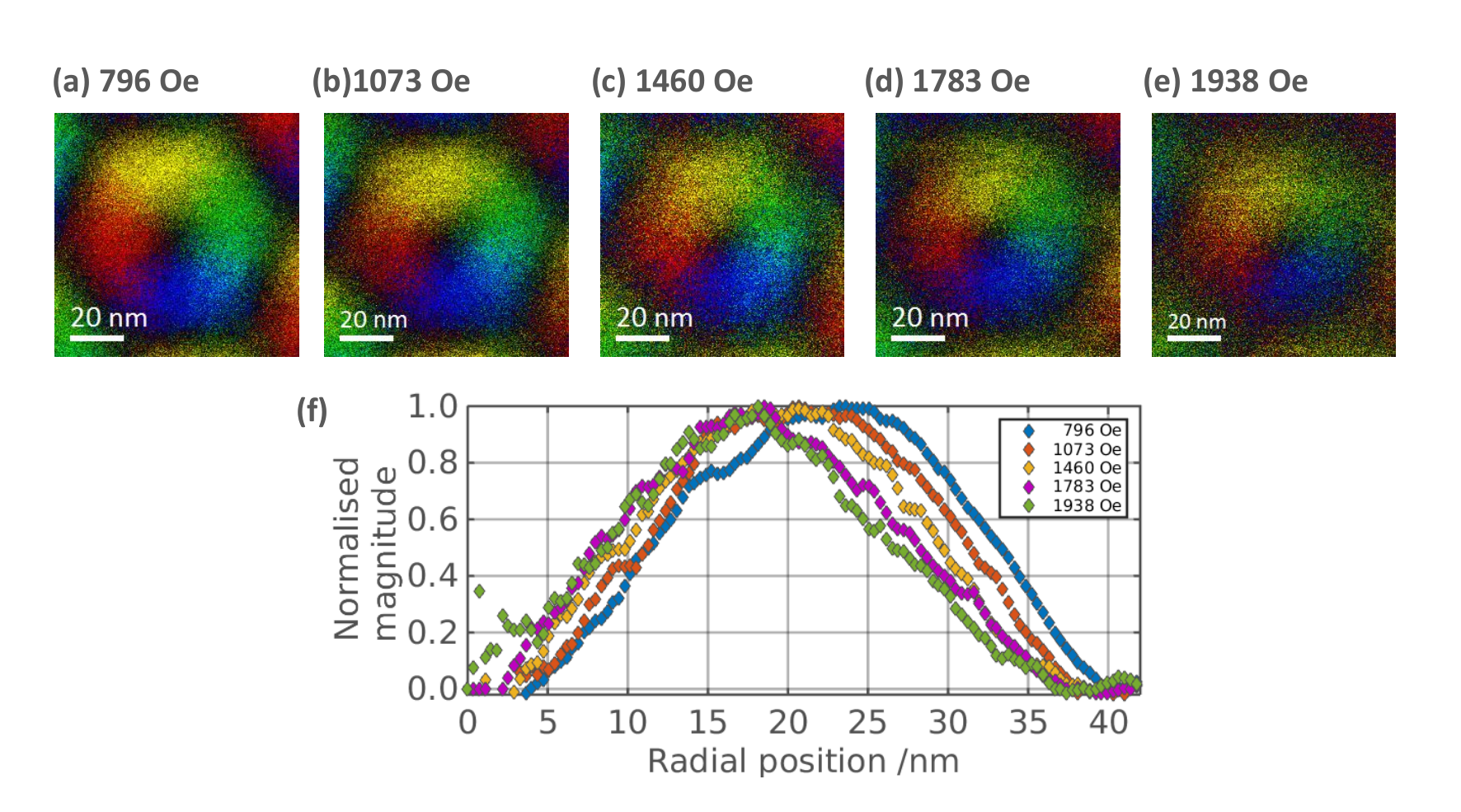}
\caption{(a)- (e) Colour in-plane vector maps of average lattice skyrmions at the applied field strengths indicated.(g) Radially averaged intensity profiles from magnitude images of lattice skyrmions with applied magnetic fields.
\label{fig:dpc_sk_profiles}}
\end{figure*} 

\begin{figure*}
\includegraphics[width=16cm]{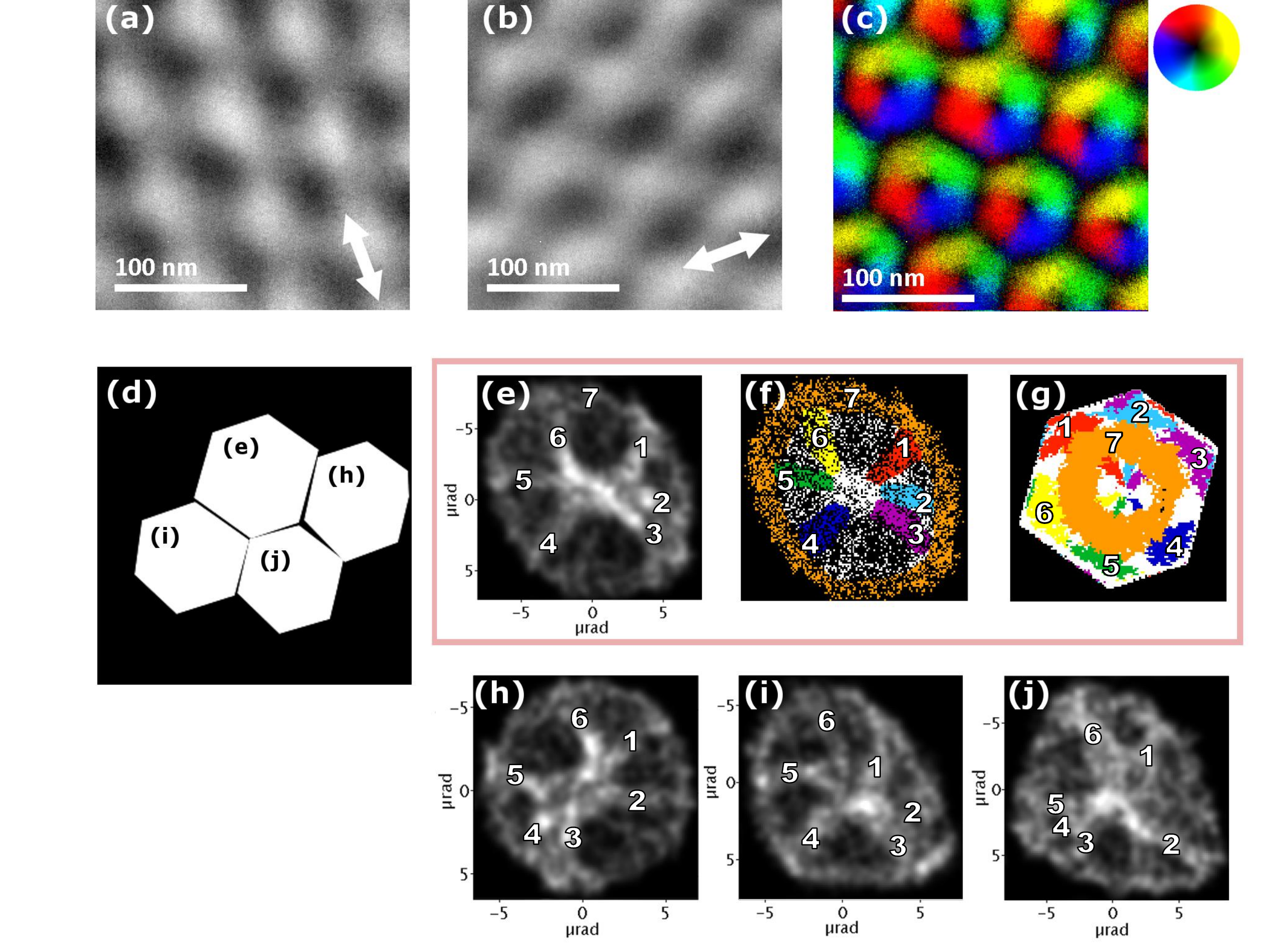}
\caption{(a-c) Orthogonal component and colour vector in-plane images produced by Pixelated DPC analysis. (d) Map showing locations for bi-variate histogram analyses of individual skyrmions. (e) bivariate histogram for a single skyrmion with symmetry spokes numbered 1-6 and outer ring numbered 7. (f) bivariate histogram from (e) but with each element 1-7 colour coded. (g) reconstructed real space map of the skyrmion (e) with colour coded zones corresponding to the elements shown in (f). (h)-(j) bivariate histograms highlighting variation in individual skyrmion structure.
\label{fig:fig_indiv_sk_variation}}
\end{figure*}

\section{Measurement of structural evolution }
The evolution of lattice skyrmion structure in response to an applied field was studied by forming average skyrmions from DPC lattice images, at field strengths from 796 to 1938 Oe, figure \ref{fig:dpc_sk_profiles}(a-e)(colour vector images of the skyrmion lattices from which the average skyrmions are produced can be viewed in figure S5 of the Supplementary Information). It can be seen that the skyrmion remained similar at all field strengths although the diameter of the core appeared to reduce with increasing field strength. This was confirmed by radially analysing the intensity profile from magnitude images, as was performed earlier in figure \ref{fig:dpc_sk_averaged}(f). Plotting profiles as a function of applied field in figure \ref{fig:dpc_sk_profiles}(f) showed that the core radius decreased by $7 nm$ ( and hence diameter = $14 nm$), evidenced by inward shifting of the peak position, from $H_{appl}=796$ to $H_{appl}=1938 Oe$. This observation is highly important since it signifies that the magnetic structure of the Bloch type skyrmions in this system is not fixed but results from energetic interplay between various terms and that the skyrmion profile is a non-linear response to this. Furthermore, that the core reduces in diameter with increasing field strength, supports the core magnetisation, which lies out of plane, is oriented anti-parallel to both the applied field and to the magnetisation at the skyrmion boundary.

\section{Variation about perfect skyrmion structure}
As our images in figure \ref{fig:dpc_explain_skyrmions} showed, skyrmions form highly ordered lattices, but on an invidual basis there appears to be a natural level of structural variation. In order to investigate this further, we employed an enhanced DPC method that we have developed where a pixelated detector (Medipix3\cite{Medipix3}, 256x256 pixels) replaced the segmented detector\cite{krajnak_pixdpc}. For each electron beam scan location on the specimen, images of the transmitted bright field electron disc were obtained. Routines for processing the resulting 4D dataset have been developed by us \cite{krajnak_pixdpc} in order to attain measurement of the Lorentz deflection of the electron disc with better than sub-pixel accuracy. Applying this method we have produced DPC component and colour in-plane vector images of lattice skyrmions for $H_{appl}= 796 Oe$  in figure \ref{fig:fig_indiv_sk_variation}(a)-(c) (STEM images 256x256 pixels, beam convergence semi-angle $\alpha = 2.2 mradians$, spatial resolution better than $1 nm$, pixel resolution $0.92nm$). As our pixelated detector directly counts single electrons with zero noise, we gain greater sensitivity and can reduce the statistical uncertainty of the disc deflection measurement to a variance of $\pm 0.03 pixels$, or $\pm0.5 \mu radians$ of Lorentz beam deflection. Thus, from figures \ref{fig:fig_indiv_sk_variation}(a)$\&$(b), bivariate histogram analyses can be performed for individual skyrmions whereas previously with the segmented detector this was only possible for the averaged lattice skyrmion in figure \ref{fig:dpc_sk_averaged}(e). In figure \ref{fig:fig_indiv_sk_variation}(d) we show a map highlighting four individual skyrmions from the lattice seen in figure  \ref{fig:fig_indiv_sk_variation}(c). A bivariate histogram for the topmost skyrmion is shown in figure  \ref{fig:fig_indiv_sk_variation}(e) where it can be seen that the histogram again consists of six symmetry spokes, numbered 1-6, and an outer ring, numbered 7. In contrast to  figure \ref{fig:dpc_sk_averaged}(e), the outer ring depicts an increased angular deflection larger than $\pm 5 \mu radians$ which we ascribe to the data here being acquired from thicker region of the nanowedge, $80-100 nm$, than that in  figure \ref{fig:dpc_sk_averaged}. In order to understand how the features of the bivariate historam correspond to real space locations within the structure of the skyrmion we have selected and colour coded these, as illustrated in figure \ref{fig:fig_indiv_sk_variation}(f), and reconstructed each in figure \ref{fig:fig_indiv_sk_variation}(g). From figure \ref{fig:fig_indiv_sk_variation}(g) it can be seen that the symmetry spokes numbered 1-6 in figure  \ref{fig:fig_indiv_sk_variation}(f) correspond to discrete zones at the corner regions of the skyrmion and also to regions which project inwards towards it's centre. The outer ring in figure  \ref{fig:fig_indiv_sk_variation}(f) numbered 7 , where the in-plane magnetic component of magnetic induction is strongest, occupies the region midway between the boundary and the skyrmion centre in figure  \ref{fig:fig_indiv_sk_variation}(g). Relating the form of the bivariate histograms to the real space features highlights that this is an effective method by which to characterise distortions. A key indicator of symmetry reduction is given by the symmetry spokes not possessing equal angular separation, which, in real space, corresponds to unequal hexagon boundary lengths. A clear example is shown when comparing figures \ref{fig:fig_indiv_sk_variation}(f) and (g). In figure \ref{fig:fig_indiv_sk_variation}(f) symmetry spokes 1,2 and 3 have a closer angular spacing  than spokes 4, 5 and 6. Examination of figure \ref{fig:fig_indiv_sk_variation}(g) reveals that the side lengths of the boundaries of the skyrmion are shorter for 1-2 and 2-3 than the other sides. Similar distortions are evident in figures  \ref{fig:fig_indiv_sk_variation}(h)-(j) and by referring back to figure  \ref{fig:fig_indiv_sk_variation}(d) for locations in figure(c) this observation can be correlated to there being non-equal hexagon boundary lengths and non-equal areal occupation of the colour domains within individual skyrmions. In the present study we were unable to correlate the observed variations in symmetry with any identifiable features in the single crystal samples although we cannot rule out that such distortions might have been caused by undetected strain or bending of the specimen.  

\section{Theoretical prediction of structure for lattice skyrmions}
\begin{figure*}
\includegraphics[width=12cm]{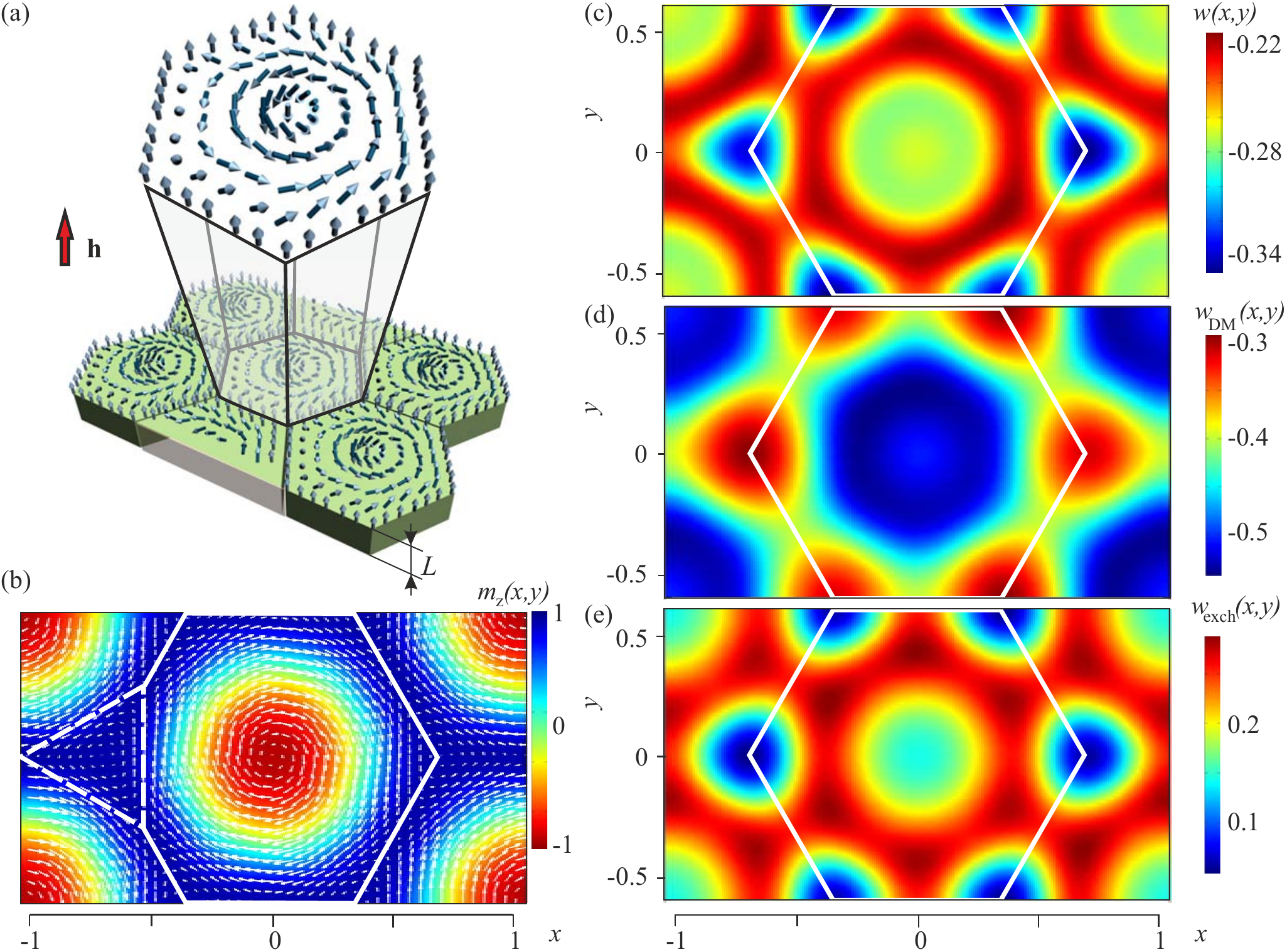}
\caption{(a) Hexagonal skyrmion lattice and the unit cell with the axisymmetric distribution of the magnetization near the
center. (b) Contour plot for the averaged over the film thickness $m_z$-component of the magnetization on the plane $(x, y)$. $H/H_D=0.1$   The white arrows show the corresponding distribution of the in-plane component of the magnetization. The hexagonal unit cell is highlighted by the white hexagon. A white triangle with the dashed contour shows the specific region formed in the corners of the hexagon and acquiring additional rotation. Distributions of the total (c), DM (d) and exchange (e) energy densities averaged over the film thickness and plotted on the plane $(x, y)$. The energy densities are expressed in units $A/D^2$.
\label{structure}}
\end{figure*}
\subsection{Phenomenological model}
The standard model for magnetic states in cubic non-centrosymmetric ferromagnets is based on the energy density functional \cite{Dz64,Bak80}
\begin{equation}
w (x,y,z) =A\,(\mathbf{grad}\,\mathbf{m})^2 + D\,\mathbf{m}\cdot \mathrm{rot}\,\mathbf{m} -\mu_0 \,M  \mathbf{m} \cdot \mathbf{H},
\label{density}
\end{equation}
with the principal interactions essential to stabilize modulated states: the exchange stiffness with constant $A>0$, 
Dzyaloshinskii-Moriya coupling energy with constant $D$, and the Zeeman energy;
$\mathbf{m}$ is the unity vector along the magnetization vector  $\mathbf{M} = \mathbf{m} M$, and $\mathbf{H}$ is the magnetic field applied along the $z-$ axis.
We investigate  the functional (\ref{density}) in a film of thickness $L$ with free boundary conditions at the film surfaces $z = \pm L/2$ and infinite in $x-$ and $y-$ directions. 
%
%
As an example, we carried out the full analysis of the rigorous solutions for hexagonal skyrmion lattices (Fig. \ref{structure} (a)) in the film with the \textit{confinement ratio} $\nu = L/L_D=1.51$ where $L_D = 4\pi A/|D|$ is the \textit{period} of the equilibrium helix in zero field (for details of the numerical methods see Refs. \cite{Keesman,LeonovThesis}). 
Due to the denser packing of individual skyrmions, such a hexagonal lattice provides smaller energy density in comparison with e.g.  square skyrmion arrangement \cite{LeonovThesis}. 
%

\subsection{Internal structure of an ideal skyrmion lattice: double twist versus compatibility}
Figure \ref{structure} (b) shows the contour plot for the $m_z(x,y)$ component of the magnetization with the in-plane components $m_x$  and $m_y$ shown by the white arrows.  
Figure \ref{structure} (c)-(e) shows the distributions of the free-energy density $w(x,y)$ (c), the rotational DM energy density $w_{DM}(x,y)$ (d), and the exchange energy density $w_{ex}(x,y)$ (e).
All distributions in Fig. \ref{structure} have been averaged over the film thickness.

\begin{figure*}
\includegraphics[width=12cm]{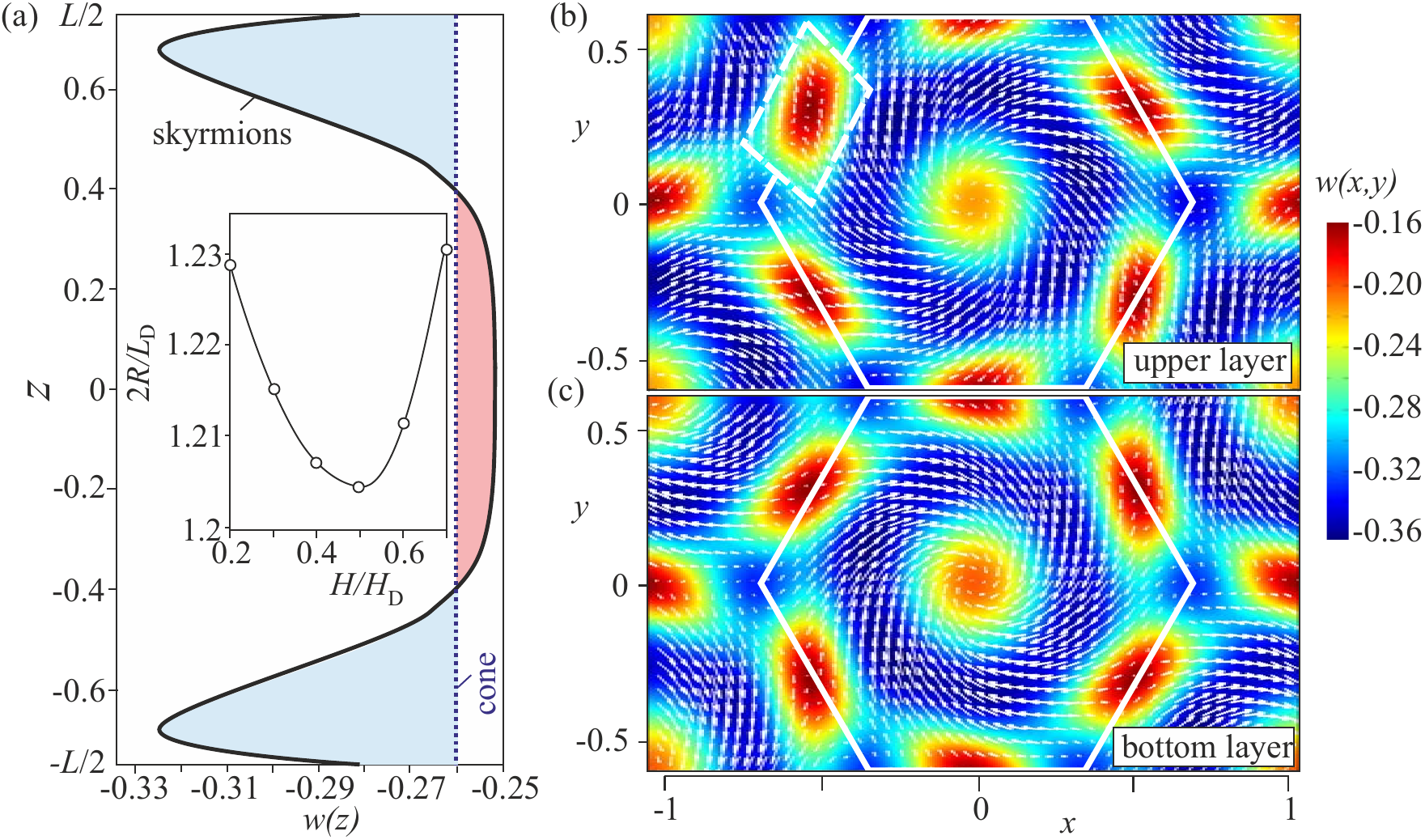}
\caption{(a) The averaged over the plane $(x,y)$  energy density $w(z)$ (black solid line)  plotted across the film thickness. The blue-shaded regions indicate the parts of the energy density $w(z)$ lower than the energy density of the conical phase (dotted blue line), whereas the red-shaded regions - the parts with the higher energy density. (b), (c) The energy density distributions $w(\pm L/2,x,y)$ for the upper and bottom surface layers. White hexagons with solid contours signify the unit cell of the skyrmion hexagonal lattice, the white rectangle with the dashed contour - specific regions with the anti-skyrmion structure formed at the inter-skyrmion areas. 
The white arrows in (b) and (c) show the corresponding distribution of the in-plane magnetization and elucidate the structure of the surface twists.
\label{twists}}
\end{figure*}

The skyrmions preserve axisymmetric distribution of the magnetization and the energy densities only near the centers of the hexagonal cells (the hexagonal cells are marked by the white-colored hexagons in Figs. \ref{structure} (b)-(e))  while the inter-skyrmion regions are strongly distorted. It is the consequence of an inherent frustration built into models with chiral couplings: the system cannot fill the whole space with the ideal, energetically most favoured double-twisted motifs.
%
%
In particular, the vertices of the hexagon  serve as sources for triangular regions of anti-skyrmions (marked by the dashed line in Fig. \ref{structure} (b)) with the magnetizations opposite to that in the center of the ”main” hexagon.
This leads to the decrease of the exchange energy density in these regions (Fig. \ref{structure} (e)) with the simultaneous growth in the rotational energy density (Fig. \ref{structure} (d)). 
The total energy density (Fig. \ref{structure} (c)), however, has benefit in the corners of the hexagons.
That is the ring with the negative energy density, which is known to form at the outskirt of an isolated metastable skyrmion within the saturated state and to prevent a skyrmion from collapse (see e.g. Fig. 2 (d) in Ref. \cite{Roessler2011}), acquires hexagonal symmetry once skyrmions are condensed into the lattice.  
The situation is opposite along the apothems of the hexagons:  the smooth rotation of the magnetization from one skyrmion center to another leads to the lower DM energy density (Fig. \ref{structure} (d)) and the larger exchange energy density (Fig. \ref{structure} (e)). 
The total energy density (Fig. \ref{structure} (c)) along the apothem of the hexagon has slightly larger value than along the diagonal.
Note the formation of rectangular-shaped regions with the anti-skyrmion structure in the inter-skyrmion area where skyrmions meet by their sides, which are in particular well  discernible for the surface energy distributions in Figs. \ref{twists} (b), (c) (see further text for details).


\subsection{Chiral surface twists and thermodynamical stability of hexagonal skyrmion lattices in nanolayers of cubic helimagnets}
In Refs. \cite{Leonov2015b} it was shown that a stabilization mechanism for hexagonal skyrmion lattices over the conical phase in thin-films of non-centrosymmetric ferromagnets is provided by the formation of so called chiral twists arising near the surfaces and decaying into the depth of the sample. 
These chiral surface twists  arise owing to the gradient terms  $m_x\partial m_y/\partial z-m_y\partial m_x/\partial z$ in the DM energy density and become evident in additional rotation of the in-plane component of the magnetization outward or inward from the center (see white arrows in Figs. \ref{twists} (b), (c) showing $m_x$ and $m_y$ components on two surfaces).

The surface twists significantly alter the energetics of the hexagonal skyrmion lattices (Fig. \ref{twists} (a)): the energy density $w(z)$ - the energy density obtained after averaging in the plane $(x,y)$ for a given coordinate $z$ - gains two pockets with the negative energy density over the conical phase. 
Therefore, the thermodynamical stability of skyrmions rests merely on the comparison of the surface areas of $w(z)$ below (blue-colored regions) and above (red-colored regions) the energy density of the conical spiral (dotted line in Fig. \ref{twists} (a)).
The magnetic phase diagram obtained for free-standing cubic helimagnet nanolayers in Ref. \cite{Leonov2015b}  shows a vast region with the thermodynamically stable hexagonal skyrmion lattices in a broad range of the confinement ratios $\nu$. 
The energy density distributions $w(\pm L/2,x,y)$ for the surface layers (Fig. \ref{twists} (b), (c)) are dominated by the rotational energy density due to the additional surface twists.
The energy density distribution of the layers in the middle of the films $w(0,x,y)$ are intact by the surface twists and look quantitatively the same as averaged distributions in Fig. \ref{structure} (a movie showing energy distributions for all layers can be found included as Supplementary Information).

\subsection{Evolution of the hexagonal skyrmion lattice in an applied magnetic field}
In general, as long as the surface twists modify the skyrmion solutions only in the direct vicinity of the surfaces and, moreover, any effect on the energy distributions (Fig. \ref{structure}) is averaged out, in what follows we consider the magnetization processes of the hexagonal skyrmion lattices for bulk helimagnets not confined by the surfaces.

\begin{figure*}
\includegraphics[width=12cm]{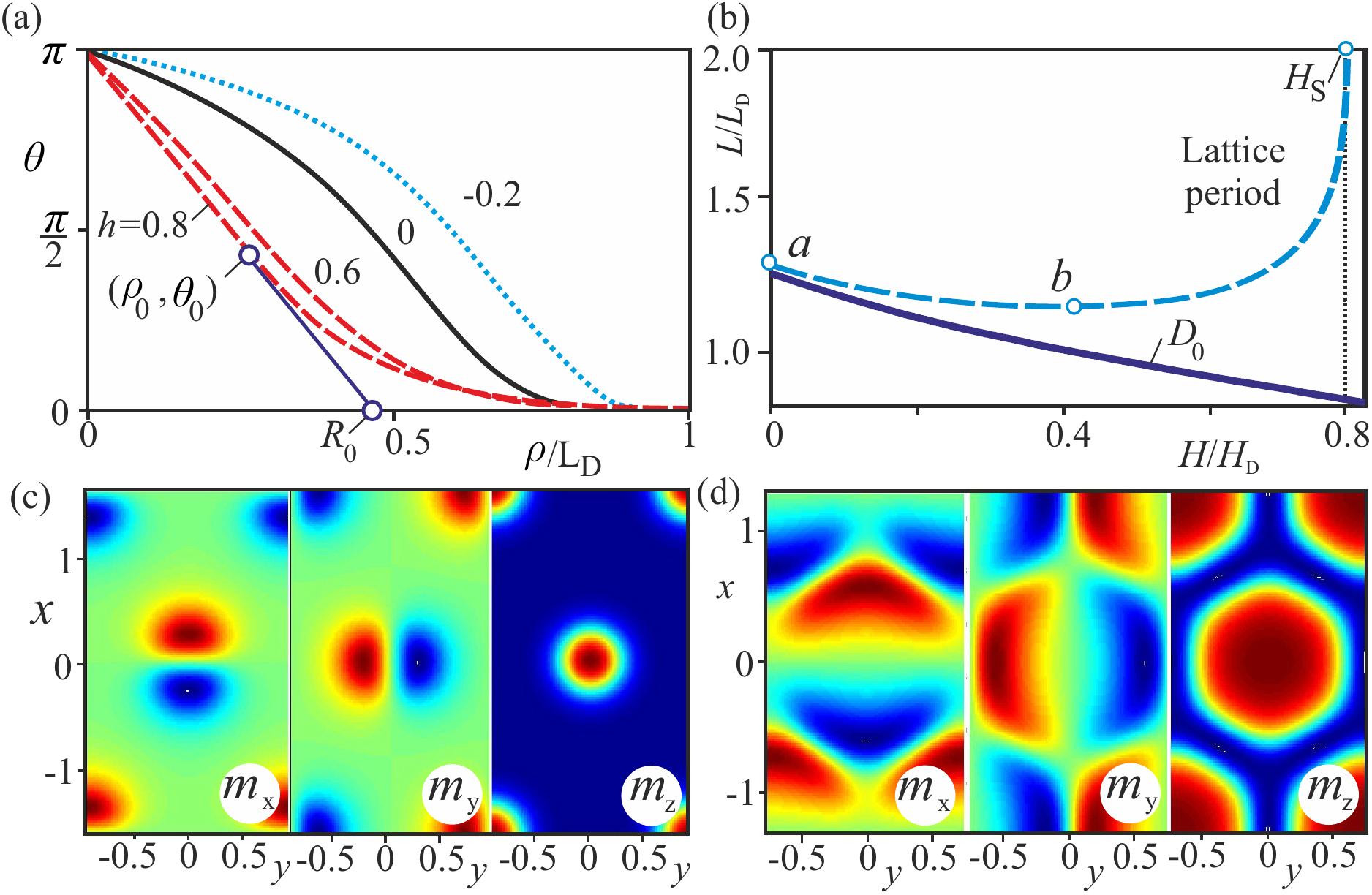}
\caption{ Evolution of the hexagonal skyrmion lattice in magnetic field applied either opposite to the magnetization in the skyrmion center (c), or parallel to it (d). The solutions are presented as angular profiles along the apothems of the hexagons (a), and contour plots for all components of the magnetization on the plane $(x, y)$:  (c) $H/H_D = 0.8$,
(d) $H/H_D = - 0.2$. (b) Equilibrium sizes of the cell core ($R_0$) and lattice period $R$ expressed in unit of $L_D$. For
positive values of the magnetic field the skyrmion lattice transforms into a system of isolated skyrmions with repulsive potential between them, whereas for negative magnetic field (not shown) it turns into the homogeneous phase through a honeycomb structure (d) with increasing lattice period.
\label{field}}
\end{figure*}

With the increasing magnetic field, the skyrmion cores are being gradually localized. 
Dashed red curves in Fig. \ref{field} (a) show angular profiles $\theta(\rho)$ as obtained along the apothem of the hexagonal unit cell.
As these profiles bear strongly localized character, a skyrmion core diameter $D_0$ can be defined in analogy to definitions for domain wall width \cite{Hubert98}, i.e. as two times the value of $R_0$, which is the coordinate of the point where the tangent at the inflection point $(\rho_0, \theta_0)$ intersects the $\rho$-axis (Fig. \ref{field} (a)).
Fig. \ref{field} (c) shows the corresponding contour plots of the skyrmion solutions for $H/H_D=0.8$. 
The lattice period, on the contrary, first shrinks reaching minimum at $H/H_D\approx 0.44$ (point $b$ in Fig. \ref{field} (b)) and then increases.
$H_D = D^2/(2A M)$ here is the \textit{saturation field} of the conical phase \cite{Bak80,JMMM94}. 
Into the homogeneous state the skyrmion lattice transforms  by infinite expansion of the period at the critical field $H_S \approx 0.8 H_D$: the skyrmion core retains a finite size (solid line in Fig. \ref{field} (b)), $D_0 (H_S) = 0.92 L_D$ and the lattice releases a set of \textit{repulsive} isolated skyrmions, owing to their topological stability.
These free skyrmions can exist far above $H_S$ up to the field of skyrmion collapse.
On decreasing the field again below $H_S$, the skyrmions can recondense into a skyrmion lattice. 
Mechanism of lattice formation through nucleation and condensation of isolated skyrmions follows a classification introduced
by DeGennes \cite{DeGennes} for (continuous) transitions into incommensurate modulated phases. For $H > H_S$ the system is magnetized homogeneously, but for $H < H_S$ the skyrmion lattice must be formed. According to De-Gennes, the fully saturated ferromagnetic state is stable locally. However for $H < H_S$, it becomes unstable with respect to certain distortions of large amplitude - skyrmions: in practice, isolated skyrmions as excitations of the saturated state nucleate near defects, and then condense into
the lattice. Such nucleation-type phase transitions are rather frequent in the condensed matter physics: (a) the
entry of magnetic flux in a type II superconductors involves nucleation of vortex lines; (b) an electric or magnetic
field induces the transition between a cholesteric and a nematic liquid crystals; (c) the magnetic samples break up into domains with increasing role of demagnetizing field \cite{DeGennes2}.
The system, however, may undergo another scenario when an enhanced coercitivity of the system prevents the formation of skyrmion lattices. In this case, skyrmions rather undergo an elliptical instability with respect to  the helical state (see Ref. \cite{Leonov2015c} for details).
Note, that $H_S$ is a function of the confinement ratio $\nu$ and strongly increases for thinner magnetic films.  
The reason for this phenomenon lies in the  surface twists that bring an additional energy into the system. 
For $\nu<0.68$ $H_S$ becomes even larger than $H_D$ \cite{Rybakov2016}. 

For a negative magnetic field applied along the magnetization in the center of skyrmion cells, both the skyrmion cores and the lattice cell size expand. Near the critical field $H_h = 0.616 H_D$ the vortex lattice consists of honeycomb-shaped cells separated from each other by narrow $360^{o}$ domain walls (Fig. \ref{field} (d)). This honey-comb lattice is highly unstable. It is hardly accessible, easily elongates into spiral state, and does not release any isolated skyrmions.

In reality, however, the hexagonal skyrmion lattices correspond to the global minimum of the magnetic energy functional (\ref{density}) only in some range of applied magnetic fields, and undergo the first-order phase transitions into other modulated phases: according to the phase diagram in Ref. \cite{Leonov2015b}, for $\nu=1.51$, hexagonal skyrmion lattices evolve from helicoids at $H/H_D \approx 0.2$ and are replaced by the cones at $H/H_D \approx 0.81$. 
These processes are accompanied by the formation of multidomain patterns of the competing phases and domain boundaries of non-trivial topology.
The periods of the hexagonal skyrmion lattices in thin films will slightly differ from those in bulk magnets. 
The inset in Fig. \ref{twists} (a) shows the inter-skyrion distance for $\nu=1.51$ in the region of the skyrmion thermodynamical stability.

\section{Discussion $\&$ Summary}
\begin{figure*}
\includegraphics[width=18cm]{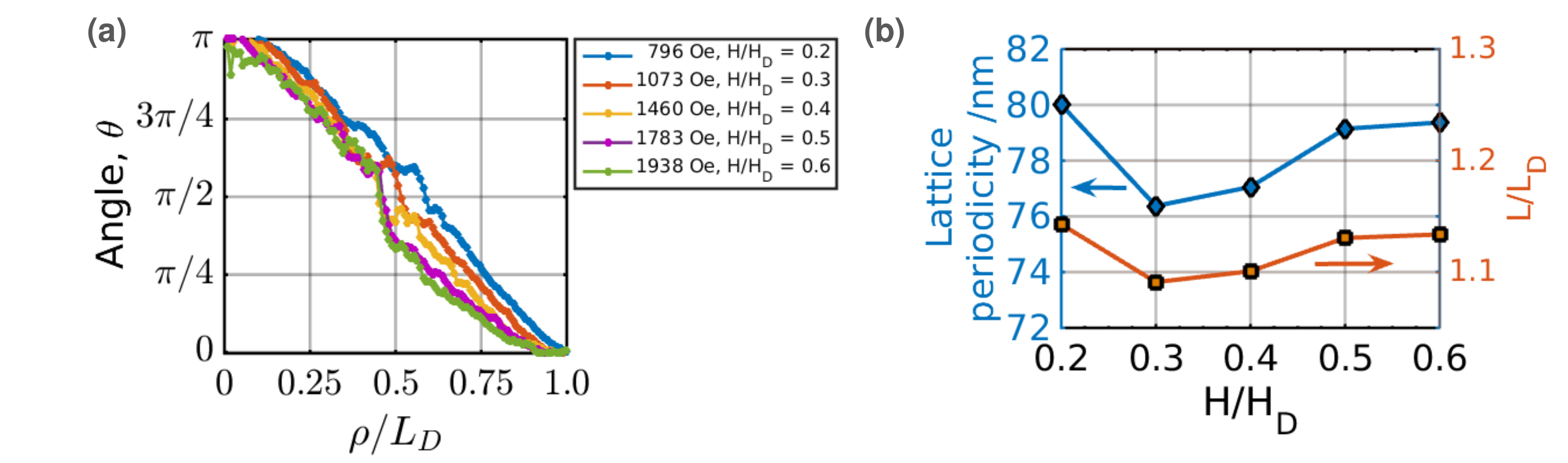}
\caption{(a) Angular profiles of lattice skyrmions from DPC experimental data plotted as a function of $L_D$. (b) Plot of the lattice periodicity variation in response to an applied field. Blue points map to the experimental length scale measured in $nm$, orange points map to lattice period expressed in units of $L_D$.
\label{fig:fig_expt_comp_theory}}
\end{figure*}

Through our combination of highly resolved experimental imaging and theoretical modelling we have revealed new insight into the internal structure and behaviour of magnetic skyrmions. We have shown that on average lattice skyrmions possess a pronounced internal six-fold symmetry and in response to an applied magnetic field of increasing strength we have measured a reduction in skyrmion core diameter. Comparison of the core size evolution with the prediction from the theoretical model,  figure \ref{field}(a), can be made by re-plotting the experimental data from figure \ref{fig:dpc_sk_profiles}(f) as skyrmion spin angle  in figure \ref{fig:fig_expt_comp_theory}(a), with the x-axis spatial coordinate expressed as a function of $L_D$. We have also experimentally measured the variation of skyrmion lattice period (employing radial profile measurements from autocorrelation of images such as figure \ref{fig:dpc_explain_skyrmions}(e) - further information given in Supplementary Information) and plot the lattice period variation in  figure \ref{fig:fig_expt_comp_theory}(b). It can be seen that across the range of applied fields that lattice period varied non-linearly in the range from 80 to 76 nm. Again our experimental observation shows good qualtitative trend agreement with the  theoretical variation depicted in figure\ref{field}(b).  

Taking together our observations and the agreement with theory, we can conclude that the magnetic response of the skyrmion lattice phase to applied fields is non-linear and governed by energetic competition between the exchange, DMI, anisotropy and magnetostatic energy terms. Furthermore, our model predicts that lattice skyrmions in a free-standing nanowedge could possess significant three-dimensional structure variation through the existence of twist states close to the material surfaces and which stabilise the skyrmion lattice phase. From the experimental results reported here we are unable to confirm the presence of surface states but we are currently conducting careful investigations to find evidence or otherwise of their existence. 

Having demonstrated the capacity of the lattice skyrmion internal structure and lattice period to be altered in response to an applied field, our observation of natural structure variation is perhaps not so surprising. That symmetry lowering distortions of the internal structure occur, suggest that these are likely in response to a source of intrinsic strain. 

Whether the source is mechanical, from strain distributions within the nanowedge crystal, or due to ``magnetic'' strain distributions in the skyrmion crystal lattice itself remains a topic for further investigation. It has been demonstrated that at crystal grain boundaries \cite{matsumoto_scienceadvances} and through the application of modest mechanical strain there can be significant influence on the skyrmion lattice structure in FeGe \cite{shibata_naturenano}. Alternatively magnetic strains causing defect structures in skyrmion lattices have also been shown to exist at skyrmion lattice rotational domain boundaries \cite{Rajeswari15}.

Finally, as an extension to this work, we propose that if isolated, single skyrmions can be obtained in the cubic B20 materials, that it is interesting to investigate their internal symmetry and precise spin profiles in order to compare with their lattice bound counterparts. It is possible to obtain unbound skyrmions under high magnetic field conditions as shown by Yu et al. \cite{Yu10} and in our own experiments (unpublished). We intend to carry out such an investigation using our DPC pixelated detection technique.

\section{Acknowledgements}
The authors would like to thank M. Miyagawa, Y. Kousaka, T. Koyama, Ts. Koyama, S. Mori, J. Akimitsu, K. Inoue for the preparation of the bulk FeGe crystal and the TEM specimen. Also, we would like to thank G. Tatara and T. Monchesky for helpful discussions.

This work was supported by funding from the following sources: JSPS Grant-in-Aid for ScientificResearch (S) (No. 25220803), JSPS Brain Circulation Project (R2507), JSPS Core-to-Core Program “A. Advanced Research Networks”,  the MEXT program for promoting the enhancement of research universities (Hiroshima University) and the UK Engineering and Physical Sciences Research Council, grant number EP/M024423/1.

\end{document}